\documentclass[a4paper, 11pt]{article}
\usepackage{slashed, color}
\usepackage{multirow}
\usepackage{tikz}
\usepackage{hyperref}
\usepackage{latexsym, amsmath, amssymb,  graphicx,  subfigure}
\usepackage{simplewick}
\usepackage{mathrsfs}
\usepackage{cancel}
\usepackage{cite}
\let\nc\newcommand
\newcommand{\Tr}{\mathop{\mathrm{Tr}}\nolimits}
\nc{\thetab}{\bar{\theta}}
\nc{\Gammab}{\wbar{\Gamma}}
\nc{\kappab}{\bar{\kappa}}
\nc{\psib}{\bar{\psi}}
\nc{\Lambdab}{\bar{\Lambda}}
\nc{\rhob}{\bar{\rho}}  
\title{Gauge and Supersymmetry Invariance of $\mathcal{N} =2$ Boundary Chern-Simons Theory}
\author{Mir Faizal$^{1, 2}$, Yuan Luo$^3$, Douglas J Smith$^4$, Meng-Chwan Tan$^3$, Qin Zhao$^3$ \\
$^1$ Department of Physics and Astronomy,
University of Lethbridge, \\
Lethbridge,   Alberta, T1K 3M4, Canada. \\ $^2$ Department of Physics and Astronomy,   University of Waterloo,  \\  Waterloo, 
Ontario N2L 3G1,   Canada. \\
$^3$ Department of Physics, National University of Singapore, \\
 2 Science Drive 3, Singapore. \\
$^4$ Department of Mathematics,  Durham University, 
\\ Durham, DH1 3LE, United Kingdom.}
\date{}
\begin{document}

\begin{flushright}
{\bf DCPT-16/03}
\end{flushright}

{\let\newpage\relax\maketitle}

\begin{tabular}{c}
\textit{E-mail:} \href{mailto:f2mir@uwaterloo.ca}{f2mir@uwaterloo.ca},
\href{mailto:phyluoy@nus.edu.sg}{phyluoy@nus.edu.sg}, \tabularnewline \href{mailto:douglas.smith@durham.ac.uk}{douglas.smith@durham.ac.uk}, \href{mailto:mctan@nus.edu.sg}{mctan@nus.edu.sg}, 
\href{mailto:zhaoqin@u.nus.edu}{zhaoqin@u.nus.edu} \tabularnewline
\end{tabular}

\begin{abstract}
In this paper, we study the restoration of gauge symmetry and up to half the supersymmetry ($\mathcal{N}=(2,0)$ or $\mathcal{N}=(1,1)$ in two dimensions) for $\mathcal{N}=2$ non-Abelian Chern-Simons theories in the presence of a boundary. We describe the boundary action which is a supersymmetric WZW model coupled to the bulk Chern-Simons theory. Unlike the $\mathcal{N}=1$ case, higher supersymmetry ($\mathcal{N}=(2,0)$) will endow the group manifold of the WZW model with a complex structure. Therefore, the $\mathcal{N}=(2,0)$ WZW model in our paper is constructed via a coset space $G_c/G$, where $G$ is the same as the gauge group in the Chern-Simons action. 
\end{abstract}
\section{Introduction}

The low energy effective action for M2-branes with manifest $\mathcal{N} = 8$ supersymmetry is described by a Chern-Simons-matter theory called BLG theory \cite{Bagger:2006sk, Gustavsson:2007vu, bagger_gauge_2008, bagger_comments_2008, gustavsson_selfdual_2008}. 
This theory is based on a Lie 3-algebra. The requirement of a finite dimensional algebra with positive definite metric restricts the application to essentially only 2 M2-branes. However, the ABJM theory
generalizes the BLG theory to 
a Chern-Simons-matter theory describing any number of M2-branes, but with only manifest $\mathcal{N} = 6$ supersymmetry. The gauge sector of this theory is described by two Chern-Simons theories 
with the gauge group $U_k (N) \times U_{-k}(N)$,  
 where $\pm k$  are the  levels for the two Chern-Simons theories  \cite{aharony_n6_2008}.
In fact for the only suitable finite dimensional 3-algebra the BLG theory can be re-written in a form which demonstrates that the
ABJM action for $N=2$ is equivalent to the BLG action \cite{VanRaamsdonk:2008ft}.

It is possible for M2-branes to end on other objects such as M5-branes, M9-branes,  
and gravitational waves. This is relevant for heterotic string theory \cite{horava_heterotic_1996, horava_eleven-dimensional_1996, Lambert:2015ita} while general boundary conditions for multiple M2-branes were studied in \cite{berman_boundary_2010}. In fact, the study of multiple M2-branes ending on M5-branes in a $C$-field background has led to the proposal of a novel quantum geometry using the 3-algebra structure \cite{chu_towards_2009}, rather different from previous attempts to describe noncommutative strings based on a single
M2-brane \cite{bergshoeff_noncommutative_2000, kawamoto_open_2000, berman_open_2004}.
Also, the BLG action with an infinite-dimensional 3-algebra has been proposed as the action for an M5-brane in a large $C$-field background \cite{ho_concise_2009}. 
So, it is important to analyze the ABJM theory in the presence of a boundary.

The  connection between three-dimensional 
Chern-Simons theories and two dimensional conformal field theories is well known \cite{witten_quantum_1989}. 
In the presence of a boundary we can
impose appropriate boundary conditions \cite{moore_tam_1989, elitzur_remarks_1989},
with the result that a component of the gauge field appears linearly in the action. 
It can thus be integrated out, imposing a constraint. A WZW model on the boundary is obtained as a solution to 
this constraint. In this way the bulk
gauge potential gets  replaced by the boundary WZW degrees of freedom.
Alternatively, it is possible to define a boundary action by requiring the gauge transformation of the boundary action to exactly cancel the boundary term generated from the gauge transformation of the bulk action \cite{chu_multiple_2010}. For ABJM theory the
matter sector is gauge invariant even in the presence of a boundary, so only the gauge invariance  of the Chern-Simons theories in presence of a boundary has to be restored.

A very similar issue is that boundary terms break the supersymmetry of a theory. 
This is because in general the supersymmetry transformations of supersymmetric actions give  rise to total derivatives, 
and hence on a manifold without a boundary, these total derivatives vanish. However, in the presence of a boundary
it is only possible to restore at most half the supersymmetry by imposing appropriate boundary conditions \cite{belyaev_boundary_2006-1, belyaev_boundary_2006}.
There are various constraints generated from supersymmetry on the possible boundary conditions \cite{van_nieuwenhuizen_consistent_2005, lindstrom_consistent_2003, di_vecchia_fermionic_1982, di_vecchia_fermionic_1983, igarashi_supersymmetry_1984}.
These boundary conditions   are imposed on-shell, and so the off-shell supersymmetry is still broken. As most supersymmetric theories are quantized using path integral
formalism, and path integral formalism  uses
off-shell fields, 
it is important to try to construct actions which preserve some supersymmetry off-shell. It is possible to restore part of the off-shell supersymmetry by  adding  
new boundary terms to the original action, such that the supersymmetric transformations of these boundary terms cancel the 
boundary pieces generated by the supersymmetric transformations of the bulk theory. 
This approach was developed by Belyaev and van Nieuwenhuizen and applied to many examples, including
three-dimensional Abelian gauge theories with $\mathcal{N} =1$ supersymmetry in the presence of a boundary \cite{belyaev_rigid_2008}. This approach has also been used to analyze
an Abelian version of the ABJM theory, in presence of a boundary,  in $\mathcal{N} = 2$ superspace  \cite{berman_membranes_2009}. 
The invariance of the full non-Abelian 
ABJM theory, in the presence of a boundary,  under both gauge   and supersymmetric transformations has only been discussed in detail in
$\mathcal{N} = 1$ superspace \cite{faizal_supersymmetric_2012}. 
The supersymmetry of a matter-Born-Infeld action, in the presence of a boundary, has been discussed in 
$\mathcal{N} = 2$ superspace \cite{faizal_non-anticommutativity_2013}.
In the application to ABJM theory
the analysis done for the supersymmetric invariance of the matter sector will follow from a similar analysis done for the gauge sector. So, in this paper, we will only analyse the gauge and supersymmetric invariance of $\mathcal{N} = 2$ Chern-Simons theories in the presence of a boundary. 
We note that the superspace boundary actions for $\mathcal{N} = 2$ non-Abelian Chern-Simons theories were presented in \cite{armoni_defects_2015} in the context of systems of D3-branes with boundaries on 5-branes in type IIB. In this paper we discuss detailed properties of such boundary actions and give expressions for the component actions. As well as giving a more explicit description of the $\mathcal{N} = 2$ systems with boundary, we expect these results will be important for attempts to generalize to higher supersymmetry where superspace formulations become less practical. One obvious motivation is to study ABJM theory with a boundary where we can expect to preserve up to 6 supercharges, either $\mathcal{N} =(4,2)$ or $\mathcal{N} =(6,0)$ corresponding to M2-branes ending on M5-branes or M9-branes respectively.

\section{Boundary Supersymmetry }
In this section we will review an approach by Belyaev and van Nieuwenhuizen to constructing supersymmetric theories on manifolds with boundaries 
\cite{belyaev_rigid_2008}, see e.g.\ \cite{berman_membranes_2009, faizal_supersymmetric_2012, faizal_non-anticommutativity_2013,armoni_defects_2015} for related applications. The key concept is
to introduce new degrees of freedom on the boundary in order to compensate for
the variations of the bulk theory which no longer vanish when there is a boundary. Since a boundary breaks some translation 
invariance, it is not possible to
preserve all supersymmetry, but half the original supersymmetry can be preserved.

In this section we will review the techniques to derive the required boundary action, first for $\mathcal{N} = 1$ and then for $\mathcal{N} = 2$ supersymmetry. We will specifically work in three spacetime dimensions with a two dimensional boundary. Following the notation in \cite{belyaev_rigid_2008} we use coordinates $x^{\mu}$ where $\mu \in \{0, 1, 3\}$ in the bulk. We take $x^3 \le 0$ so we have a boundary at $x^3 = 0$ with boundary coordinates $x^m$ where $m \in \{0, 1\}$, and $\int d^3x \partial_3 X = \int d^2x X$.

\subsection{\texorpdfstring{$\mathcal{N}=1$}{N=1} Supersymmetry}
Let us first define our notation and review the minimal case of $\mathcal{N} = 1$ supersymmetry. We first introduce
the spinor $\theta_{\alpha}$ as two component anti-commuting
parameters with odd Grassmann parity and let $\theta^{2}=\frac{1}{2}\theta^{\alpha}\theta_{\alpha}=-\frac{1}{2}\theta_{\alpha}C^{\alpha\beta}\theta_{\beta}$, where $C^{\alpha\beta}$ is an anti-symmetric tensor used to raise
	and lower spinor indices  and $C_{\alpha\beta}C^{\gamma\sigma}=\delta^\gamma_{[\alpha}\delta^\sigma_{\beta]}$ (see appendix \ref{convention}). 
The generators of $\mathcal{N} =1$ supersymmetry in three dimensions can be represented by 
$Q_{\alpha} = \partial_{\alpha}  - (\gamma^\mu \theta)_{\alpha} \partial_\mu$.
These generators of supersymmetry satisfy $\{Q_{\alpha}, Q_{\beta}\} = 2 (\gamma^\mu \partial_\mu)_{\alpha\beta}$.  We can also construct 
super-derivatives $D_{\alpha} = \partial_{\alpha}  + (\gamma^\mu \theta)_{\alpha} \partial_\mu$, 
such that  $D_{\alpha}$ commutes with the generators of supersymmetry,  
$\{D_{\alpha}, Q_{\beta}\} =0$. 
 The generators of $\mathcal{N} =1$ supersymmetry in the bulk can be decomposed into $Q_{\pm} = P_{\pm}Q$ where $P_{\pm}=  ( 1 \pm \gamma^3)/2 $ so that $\epsilon^{\alpha} Q_{\alpha} = \epsilon  P_-  Q
+  \epsilon  P_+ Q  
= \epsilon^{-}    Q_{-} + \epsilon^{+}  Q_{+}$.
The super-derivatives in the bulk can be similarly decomposed as $ D_{\alpha} =
   D_{-\alpha} +  D_{+\alpha}$. 
The bulk supercharges satisfy, 
 \[ \begin{array}{lll} \{Q_{+\alpha},Q_{+\beta}\}=2(\partial_{++})_{\alpha\beta}, &  & \{Q_{-\alpha},Q_{-\beta}\}=2(\partial_{--})_{\alpha\beta},\\ \{Q_{+\alpha},Q_{-\beta}\}=2(P_{+})_{\alpha\beta}\partial_{3}, \end{array} \]
where $(\partial_{++})_{\alpha\beta}=(P_{+}\gamma^{m})_{\alpha\beta}\partial_{m}$
and $(\partial_{--})_{\alpha\beta}=(P_{-}\gamma^{m})_{\alpha\beta}\partial_{m}$
, and
 \[ \begin{array}{lll} \{D_{+\alpha},D_{+\beta}\}=-2(\partial_{++})_{\alpha\beta}, &  & \{D_{-\alpha},D_{-\beta}\}=-2(\partial_{--})_{\alpha\beta},\\ \{D_{+\alpha},D_{-\beta}\}=-2(P_{+})_{\alpha\beta}\partial_{3}. \end{array} \]

Now, let us discuss a general $\mathcal{N} = 1$ supersymmetric invariant action in the presence of a boundary \cite{belyaev_rigid_2008}. The bulk action of $\mathcal{N} = 1$ supersymmetry is 
\begin{equation}
 S = \int d^3x \int d^2 \theta \Phi, 
 \label{n1}
\end{equation}
 where  $\Phi=a+\theta\psi -\theta^2 f$ 
 and the supersymmetry transformations are 
\begin{eqnarray}
 \delta a &=& \epsilon \psi, \nonumber \\
\delta \psi &=&  -\epsilon f + (\gamma^\mu\epsilon) \partial_\mu a, \nonumber \\
\delta f &=& -\epsilon (\gamma^\mu \partial_\mu)\psi.
\end{eqnarray}
Thus, under these supersymmetric transformations generated by $Q_{\alpha}$, the action transforms
as $\delta S= - \int d^3x\partial_\mu (\epsilon\gamma^\mu\psi )$. So, the action (\ref{n1}) is invariant under the supersymmetric 
transformations generated by $Q_{\alpha}$, in the absence of boundary. 
However, in the presence of a boundary, the  supersymmetric 
transformations generated by $Q_{\alpha}$ produce a boundary term. Thus, if we assume that a boundary exists at $x_3 =0$, 
then  the supersymmetric transformations of
the action
$\delta S = -\int d^3 x\partial_3 ( \epsilon \gamma^3  \psi) $
will generate  a boundary term. 
This breaks the supersymmetry of the resulting theory. 

We can preserve half of the supersymmetry of the resulting theory by either adding or subtracting a boundary term to the 
original action. Now if  $S_b = -\int d^3x \partial_3 \Phi|_{\theta =0}$
is the boundary term added or subtracted from the bulk action with $\mathcal{N} =1$ supersymmetry, then we have 
\begin{equation}
 \delta ( S\pm S_b) =  \mp 2  \int d^3x\partial_3 \epsilon^{\pm}\psi_{\pm}.
\end{equation} 
Hence, the action $ S^{\mp}=S\pm S_b = \int d^3x \left( (\int d^2\theta \mp \partial_3) \Phi \right) \vert_{\theta =0}$  preserves the 
supersymmetry generated by 
$\epsilon^{\mp} Q_{\mp}$, which is only half of the $\mathcal{N}=1$ supersymmetry. It is not possible to simultaneously preserve both of the supersymmetries generated by  $\epsilon^{+} Q_{+}$ and 
$\epsilon^{-} Q_{-}$, 
in the presence of a boundary.

As described in \cite{belyaev_rigid_2008}, one can construct boundary superfields by projecting bulk superfields onto the boundary.
In this paper, we denote the induced value of bulk quantities on the  boundary  by putting a prime $(')$ on them.
For example, ${\Phi}'$ is a boundary superfield derived from the bulk superfield $\Phi$.
Similarly, the boundary supercharges will be denoted by $Q_{\pm}' = \partial_{\pm} - (\gamma^m \theta)_{\pm}\partial_m $.
Note that the bulk and boundary supercharges are related by
 $Q_{\pm}' = Q_{\pm} \pm \theta_{\pm} \partial_3$, so we can write 
 $Q_{\pm}' = M_{\pm}^{-1} Q_{\pm} M_{\pm}$, where $M_{\pm} = \exp ( \pm \theta^{-} \theta_{-} \partial_3)$. The bulk super-derivatives are similarly related to the boundary super-derivatives as follows, 
 $D_{\pm}' = D_{\pm} \mp \theta_{\pm} \partial_3 = \partial_{ \pm} + (\gamma^m \theta)_{\pm}\partial_m $, and so
 $D_{\pm}' = M_{\mp}^{-1} D_{ \pm} M_{ \mp}$. 
 Now we  can write the $\mathcal{N} =1$ boundary superfields explicitly in terms of the bulk superfields, e.g. 
$\Phi_{\pm} ' = M_{\pm}^{-1} \Phi|_{\theta^\mp=0}$ defines a boundary superfield  $\Phi_{\pm} '  = a' + \theta^\pm \psi_{\pm} '$ 
from the bulk superfield $\Phi$. Note that this is arranged so that the boundary supersymmetry transformation is induced by the bulk supersymmetry transformation. Indeed it is easily seen that
we have $\delta' \Phi_{\pm}' = M_{\pm}^{-1}  \delta \Phi  $.

In the action, $\int d^2\theta$ can be replaced by $D^2$ along with restriction to $\theta = 0$. According to the commutation relations
\begin{equation}
 D^2 \mp \partial_3 =\pm i D_\mp D_\pm. 
\end{equation} 
Therefore, we can also write $S^\pm$ as:
 \begin{equation}
\begin{array}{ll}
S^\mp & =\pm i \int d^3 x D_\mp D_\pm \Phi|_{\theta=0} \\
&=\pm i \int d^3 x (D'_\mp\psi_{\pm} ')|_{\theta^\mp=0},
\end{array}
\end{equation}
where  $\psi_{\pm} '=(D_\pm \Phi)|_{\theta^\pm =0}$ is a boundary superfield.

\subsection{\texorpdfstring{$\mathcal{N} = 2$}{N=2} Supersymmetry}
Now, we can extend the method for $\mathcal{N}=1$ supersymmetry to $\mathcal{N} = 2$ supersymmetry. To this end, we first expand the $\mathcal{N} = 2$ superspace Grassmann coordinates $\theta^\alpha$, $\thetab^\alpha$ in terms of real $\mathcal{N} = 1$ coordinates $\theta^\alpha_1$, $\theta_2^\alpha$, as well as the generators and the spinor derivatives:
\begin{equation}
\begin{array}{ll}
\theta^\alpha=2^{-1/2}(\theta^\alpha_1+i\theta^\alpha_2),&\quad \thetab^\alpha=2^{-1/2}(\theta^\alpha_1-i\theta^\alpha_2),\\
Q_\alpha =2^{-1/2}(Q^1_\alpha -i Q^2_\alpha),&\quad \bar{Q}_\alpha=2^{-1/2}(Q^1_\alpha +i Q^2_\alpha),\\
D_\alpha =2^{-1/2}(D^1_\alpha -iD^2_\alpha),&\quad \bar{D}_\alpha=2^{-1/2}(D^1_\alpha +iD^2_\alpha).
\end{array}
\end{equation}
Now $Q^1_\alpha$ commute with $Q^2_\beta$, and $Q^i_\alpha$ and $Q^j_\beta$ have the same commutation relationships as those of the supercharges in $\mathcal{N} = 1$ supersymmetry, as do the $D^n_\alpha$. The super-derivatives $D^i_{\alpha}$ commute with the generators of the supersymmetry $Q^j_{\beta}$,  
$\{D^i_{\alpha}, Q^j_{\beta}\} =0$. 

We can now expand a general $\mathcal{N} = 2$ supersymmetric action into $\mathcal{N}=1$ superfields: 
\begin{equation}
\begin{array}{ll}
 S &=-\int d^3 x \int d^2\theta d^2\thetab \Phi (\theta, \thetab)|_{\theta= \thetab =0} \\
  &=\int d^3x \int d^2\theta_1 d^2\theta_2 \Phi(\theta_1,\theta_2)|_{\theta_1=\theta_2=0}.
\end{array}
\end{equation}
The $\mathcal{N} =2$ superfield $\Phi(\theta_1,\theta_2)$ can be decomposed as 
\begin{eqnarray}
  \Phi (\theta_1,\theta_2)&=& a_1 ( \theta_1) + \psi_1 (\theta_1) \theta_2 - f_1 (\theta_1) \theta_2^2 \nonumber \\ &=& 
 a_2 ( \theta_2) + \psi_2 (\theta_2) \theta_1 - f_2 (\theta_2) \theta_1^2,
\end{eqnarray}
where $a_1 ( \theta_1), a_2 ( \theta_2), \psi_1 (\theta_1), \psi_2 (\theta_2), f_1 
(\theta_1), f_2 (\theta_2) $ are $\mathcal{N} =1$ superfields
in their own right. 

According to the previous subsection, we know that, in the presence of a boundary, to restore  $\epsilon^{1\mp} Q_{1\mp}$ or $\epsilon^{2\mp} Q_{2\mp}$ supersymmetry, we need to add a boundary term $\mp\partial_3\int d^2\theta_{2} \Phi$ or $\mp\partial_3 \int d^2\theta_{1} \Phi$. A general action which recovers half of the $\mathcal{N} = 2$ supersymmetry is
\begin{equation}
\label{2boundary}
 S^{1\mp 2\mp} = \int d^3 x \left( \int d^2\theta_1 \mp \partial_3 \right) \left( \int d^2\theta_2 \mp \partial_3 \right) \Phi (\theta_1, \theta_2) \Big\vert_{\theta_1 = \theta_2 =0}.
\end{equation} 
If we preserve the supersymmetry corresponding to $\epsilon^{1\mp} Q_{1\mp}$  and $\epsilon^{2\mp} Q_{2\mp}$, then  
we will obtain a boundary theory with  $\mathcal{N} = (2, 0)$ or $\mathcal{N} = (0, 2)$ supersymmetry. However, if we preserve the supersymmetry corresponding to
$\epsilon^{1\pm} Q_{1\pm}$ and $\epsilon^{2\mp} Q_{2\mp}$, we will obtain  a boundary theory with $\mathcal{N} = (1, 1)$ supersymmetry. 

We can also describe (\ref{2boundary}) in terms of boundary fields. The boundary supercharges will be denoted by $Q_{i \pm}' = \partial_{i \pm} - (\gamma^m \theta_i)_{\pm}\partial_m $. Again we can relate the projection of the bulk supercharges to the generators of supersymmetry on the boundary as 
 $Q_{i\pm}' = Q_{i\pm} \pm \theta_{i\pm} \partial_3$, so we can write 
 $Q_{i\pm}' = M_{i\pm}^{-1} Q_{i \pm} M_{i \pm}$, where $M_{i\pm} = \exp (\pm \theta^{i-} \theta_{i-} \partial_3)$.

The bulk super-derivatives are related to the boundary super-derivatives as follows, 
 $D_{i\pm} = D_{i\pm} ' \pm \theta_{i\pm} \partial_3$, where 
 $D_{i \pm}' = \partial_{i \pm} + {(\gamma^m \theta_{i})}_{\pm}\partial_m $, are the boundary super-derivatives and so, 
 $D_{i\pm}' = M_{i\mp}^{-1} D_{i \pm} M_{i \mp}$.

We get  $ {(D^2_1 \mp \partial_3) (D^2_2 \mp \partial_3) =  -D_{1\mp}D_{1\pm}D_{2\mp}D_{2\pm}}$. So, the action which preserves the supersymmetry 
generated by $\epsilon_{1}^\mp Q_{1\mp}$ and $\epsilon_{2}^\mp Q_{2\mp}$ can also be written as 
\begin{equation}
\begin{array}{l}
 {
 S^{1\mp2\mp} =  -\int d^3 x D_{1\mp}D_{1\pm}D_{2\mp}D_{2\pm} \Phi (\theta_1, \theta_2) \big\vert_{\theta_1 = \theta_2 =0}.
}
\end{array}
\end{equation}
Furthermore, we can write this action in terms of boundary fields as 
\begin{equation}
\begin{array}{l}
{
 S^{1\mp 2\mp} = -\int d^3x D'_{1\mp} D'_{2\mp} (\Phi'_{2 \pm 1\pm} (\theta_{1\mp}, \theta_{2\mp})) \big\vert_{\theta_{1}^\mp = \theta_{2}^\mp =0},  
} 
\end{array}
\end{equation} 
where {$\Phi'_{2 \pm 1\pm} = D_{2\pm} D_{1\pm} \Phi(\theta_1 , \theta_2) \big\vert_{\theta_{1}^\pm = \theta_{2}^\pm =0}$} is a boundary superfield.  One may note that the action above can only describe  $(2,0)$ or $(0,2)$ supersymmetry, but it is easy to write down $S^{1\pm 2\mp}$ in a similar way.

\section{Chern-Simons Theory}
In this section we will apply the discussion in the previous section to explicitly describe super Chern-Simons theories in the presence of a boundary. In addition to restoring some supersymmetry, we also have to consider the effect of the boundary on the gauge symmetry. We will show that restoring this gauge symmetry is possible and that doing so leads to a WZW model on the boundary, coupled to the Chern-Simons theory in the bulk.
\subsection{\texorpdfstring{$\mathcal{N} = 1$}{N=1} Chern-Simons theory}
Now, let us review and discuss $\mathcal{N}=1$ Chern-Simons theories with boundaries \cite{belyaev_rigid_2008, berman_membranes_2009, faizal_supersymmetric_2012}.
\subsubsection*{$\mathcal{N}=1$ Abelian Chern-Simons theories}
For simplicity, we first discuss Abelian Chern-Simons theories. The action of an $\mathcal{N} = 1$ Abelian Chern-Simons theory with gauge group $H$ (with an implicit trace) is \cite{gates_jr_superspace_1983, brooks_topological_1989}
\begin{equation}
\label{1ACS}
S^1_{A}=\int d^3 x d^2\theta \Gamma^\alpha \omega_\alpha,
\end{equation}
where $\Gamma$ is a spinor superfield, with components
$$\Gamma=\chi-\theta M +(\gamma^\mu\theta) v_\mu -\theta^2[\lambda+ (\gamma^{\mu}\partial_\mu \chi)]$$
and $\omega_\alpha=D^\beta D_\alpha \Gamma_\beta$ is a super-covariant field strength. The gauge transformation is 
\begin{equation}
\label{gaugetr} 
\delta_g \Gamma_\alpha = D_\alpha \Phi \Rightarrow \left\lbrace
\begin{array}{l}
\delta_g \chi=\psi, \\
\delta_g M=f, \\
\delta_g v_\mu=\partial_\mu a,\\
\delta_g \lambda=0,
\end{array}\right.
\end{equation} 
i.e.\ $\delta_g(\chi, M, v_\mu, \lambda) =(\psi,f,\partial_\mu a, 0)$. The action (\ref{1ACS}) is only supersymmetric and gauge invariant up to a boundary term. To restore half of the supersymmetry, according to the previous section, we can add a boundary term to the bulk action, i.e.
\begin{equation}
\label{chirality}
\begin{array}{ll}
S^{1\mp}_{A} &=\int d^3 x (\int d^2\theta \mp \partial_3) [\Gamma^\alpha \omega_\alpha] \big\vert_{\theta=0} \\
&=\int d^3x \left( \lambda\lambda {-}4\epsilon^{\mu\nu\rho}v_\mu \partial_\nu v_\rho \mp \partial_3 (2\chi^{\pm}\lambda_{\pm}) \right).
\end{array}
\end{equation}

In the following discussion, without loss of generality, let us just consider $S^{1-}_A$, as the manipulation of $S^{1+}_A$ can be done in a similar manner with the opposite chirality.  In order to simplify the discussion,
we can introduce another $(1,0)$ supersymmetric invariant boundary term $S^{1-}_{sb}$ \cite{berman_membranes_2009} to cancel the coupling term of the non-propagating gaugino ---
\begin{equation}
\begin{array}{ll}
S^{1-}_{sb}&= -2i \int d^3 x\partial_3 \int d \theta^+\Gamma_-(D'_+ \Gamma_+)|_{\theta_-=0} \\
&= -2\int d^3 x\partial_3 \int d \theta^{+} \gamma^{m} \hat{\Gamma}_{-} \hat{\Sigma}^+_m \\
&=2\int d^3x\partial_3(v_m v^m + \chi^+\lambda_+ +\chi^{+}\gamma^m\partial_m\chi_{-}),
\end{array}
\end{equation}
where $\hat{\Gamma}_-$ and $\hat{\Sigma}^+_m$ are $(1,0)$ superfields \ref{appendix2}, with
\begin{equation}
\begin{array}{l}
\hat{\Gamma}_{-} =\chi_{-}+(\gamma^m\theta)_{-} v_m,\\
\hat{\Sigma}^+_m=v_m+\theta^-(\frac{1}{2}\gamma_m\lambda_+ +\partial_m\chi_-).
\end{array}
\end{equation}
Here, we mainly followed the discussion in \cite{berman_membranes_2009}, and see also \cite{belyaev_rigid_2008}. Actually, this boundary term is not necessary for the discussion of restoration of supersymmetry and gauge symmetry\footnote{However, if the action is not further modified, it is required to ensure that the boundary Euler-Lagrange equations produce standard consistent boundary conditions, without overconstraining the system \cite{belyaev_rigid_2008}.}. But, in the following, one can see that this  term can naturally give us the kinetic term of the boundary action and can give us useful hints for finding the boundary action of the $\mathcal{N}=2$ non-Abelian Chern-Simons action. In \cite{chu_towards_2009,faizal_supersymmetric_2012}, without introducing such a term, one can add the kinetic term of the boundary action by hand since it is supersymmetric invariant and gauge invariant. 

Then the action becomes
\begin{equation}
S^{1-}_A=\int d^3x \left( \lambda\lambda - 4\epsilon^{\mu\nu\rho}v_\mu \partial_\nu v_\rho + 2\partial_3 (\chi^{+}\gamma^m\partial_m\chi_{-}+v_m v^m) \right) .
\end{equation}
The variation of the action under the gauge transformation is
\begin{equation*}
\begin{array}{ll}
\delta_g S^{1-}_A &=\int d^3x 4  \partial_3 (\psi^{+}\gamma^m\partial_m\chi_{-}+\partial_{++} a v_{--} ).
\end{array}
\end{equation*}
 Here, note that $\partial_{\pm\pm}=(\gamma^m\partial_m)_{\pm}^\mp=\pm \partial_0 +\partial_1$ and $v_{\pm\pm}=\pm v_0 + v_1$.
So, this action already partially restored the gauge symmetry, namely
\begin{equation}\label{restored_gauge}
\begin{array}{ll}
\delta_g v_{--}=\partial_{--}a, & \delta_g v =\partial_3 a, \\
\delta_g M = f, & \delta_g \chi_+ = \psi_+.
\end{array}
\end{equation}
Therefore, one can set $\chi_+$ and $M$ to be zero, since we can always choose specific $f$ and $\psi_+$. However, the gauge transformation for $\chi_-$ and $v_{++}$ will still break the gauge symmetry of the action, and the gauge transformation of $v_{++}$ is usually related to those of $v_{--}$ and $v_3$ (\ref{gaugetr}). To solve this problem, it is possible to couple this theory to another boundary theory, such that the total action is gauge invariant. So let us consider a boundary term:
\begin{equation}
S^{1-}_{A,b} =S^{1-}_A(\Gamma^{g})-S^{1-}_A(\Gamma),
\end{equation}
where $\Gamma^g = ig(D'_- -i\Gamma) g^{-1}$ denotes the gauge transformation of $\Gamma$ by the scalar superfield $g=\exp (i\Phi')$, where $\Phi' = a +\theta^- \psi_-$ is a $(1,0)$ scalar superfield. Note that its bosonic component is a group element of the gauge group.
 Viewing $S^{1-}_A(\Gamma^{g})$ as a general gauge transformation by $g$, the term $S^{1-}_{A,b}$ should indeed be a boundary term, since $S^{1-}_A(\Gamma)$ should be gauge invariant in the absence of the boundary. Now, the total action $S^{1-}_A(\Gamma)+S^{1-}_{A,b}$ will clearly be gauge invariant if we choose $\Gamma^{g}$ to be gauge invariant, and this can be realized by defining the gauge transformation of $g$ as
\begin{equation}
g \rightarrow g u^{-1} \; , \;\; \Gamma \rightarrow \Gamma^{u} .
\end{equation}

An easy way to understand this boundary term is to consider $\Gamma=0$; then there is no contribution from the bulk action.
We have $\Gamma^g=-i(D'_-g) g^{-1}=D'_-\Phi'$. Note that the gauge transformation parameter $g$ is a $(1,0)$ superfield, which only leads to nonzero gauge transformations of $\chi_-$ and $v_m$. So, $\Gamma^g=D_- \Phi =\delta_g (\chi_-, v_m)=(\psi_-, \partial_m a)$. Combined with the gauge transformation (\ref{restored_gauge}), $\Gamma^g=(\psi_-, \partial_\mu a)$
we can obtain:
\begin{equation}
\begin{array}{ll}
S^{1-}_{A,b}  &=2\int d^3x \partial_3(i\psi_{-}\partial_{++}\psi_{-}+\partial_{++} a\partial_{--} a)\\
&=-2i\int d^2x d\theta_{-} (  \partial_{++}\Phi D'_{-}\Phi').
\end{array}
\end{equation}
Obviously, this is a standard $(1,0)$ supersymmetric action. 
It is easy to check that the action is invariant under the $(1,0)$ supersymmetric transformation:
\begin{equation}
\begin{array}{l}
\delta_- a = \epsilon^-\psi_-, \\ \delta_-\psi_- = \partial_{--}a \epsilon_+.
\end{array}
\end{equation}

Moreover, one can write the action as a $(1,0)$ Abelian WZW model, since $\Phi'$ corresponds to the group element $g=\exp(i\Phi')$ with $g_b \in H$:
\begin{equation}
\begin{array}{ll}
S^{1-}_{A,b} 
&=2i\int d^2x   d\theta_{-} (  \partial_{++}g g^{-1} D'_{-}g g^{-1}).
\end{array}
\end{equation}

We only considered the special case $\Gamma=0$ in the above situation, which means we restricted the gauge field $v=0$ and there is no gauge symmetry. Actually, we can restore the gauge field by replacing the partial derivative $\partial_{++}$ with $\mathcal{D}_{++}=\partial_{++} -iv_{++}$ and $D'_-$ with super-covariant derivative $\nabla'_- =D'_- -i\Gamma_-$. Also, replacing  $\int d\theta_-$ with $\nabla'_-$, we have the final form of the boundary action,
\begin{equation}
S^{1-}_{A,gWZW}= 2i\int d^2x\nabla'_-[ (g^{-1}\mathcal{D}_{++} g)(g^{-1}\nabla'_- g)]|_{\theta^-=0}.
\end{equation}

\subsubsection*{\texorpdfstring{$\mathcal{N}=1$}{N=1} non-Abelian Chern-Simons theories}
Now, one can consider non-Abelian Chern-Simons theories. The natural guess of the boundary term is a non-Abelian WZW model, which turns out to be just the case. The non-Abelian Chern-Simons action with group $G$ \cite{gates_jr_superspace_1983} is 
\begin{equation}
S^1_{nA} =\int d^3 x d^2\theta \Gamma^\alpha \Omega_\alpha, 
\end{equation}
where $\Omega_\alpha = W_\alpha-\frac{1}{3}[\Gamma^\beta,\Gamma_{\alpha\beta}]$, with  $W_{\alpha}=D^{\beta}D_{\alpha}\Gamma_{\beta}-i[\Gamma^{\beta},D_{\beta}\Gamma_{\alpha}]-\frac{{1}}{3}[\Gamma^{\beta},\{\Gamma_{\beta},\Gamma_{\alpha}\}]$
being the super-covariant field strength and  $\Gamma_{\alpha\beta}=-\frac{i}{2}[D_{(\alpha}\Gamma_{\beta)}-i\{\Gamma_{\beta},\Gamma_{\alpha}\}]$.
Then the action which preserves the half of supersymmetry $\epsilon^- Q_-$ is 
\begin{equation}
\begin{array}{ll}
\label{2halfsusy}
S^{1-}_{nA} & =\int d^3 x (d^2\theta -\partial_3) (\Gamma^\alpha \Omega_\alpha)\\
&=\int d^3 x [-4\epsilon^{\mu\nu\rho}(v_\mu \partial_\nu v_\rho  -\frac{2i}{3}v_\mu v_\nu v_\rho)+ \lambda^\alpha \lambda_\alpha  \\ &- 2\partial_3(\chi^+\lambda_{+} -\frac{2i}{3} \chi(\gamma^\mu v_\mu)\chi)]. 
\end{array}
\end{equation}

We can also introduce a supersymmetry invariant boundary term to kill the gaugino coupling term in the boundary as we did in the Abelian Chern-Simons case. One may note that the last term in (\ref{2halfsusy}) can be canceled by a half supersymmetric boundary term $\sim \int d\theta^+\gamma^m\hat{\Gamma}_- \hat{\Gamma}^+ \gamma_m \hat{\Gamma}_-$ and one may note that we set $\chi_+$ to be zero because of the similar reason in Abelian case (\ref{restored_gauge}).\footnote{From the $(1,0)$ superfield  $\hat{\Gamma}_+$ in \eqref{boundary_super} one can see $\delta_{\epsilon^-Q_-}\chi_+ = i\epsilon^-(-M+v_3)$. Since $M$ is absent in the action and the gauge symmetry of $M$ is naturally hold, one can always compensate $M$ by a gauge transformation to set the variation of $\chi_+$ zero. Hence, we can choose $\chi_+$ to be zero and keep the value unchanged under the supersymmetric variation.} Then the action becomes
\begin{equation}
\begin{array}{ll}
S^{1-}_{nA}
&=\int d^3 x [-4\epsilon^{\mu\nu\rho}(v_\mu \partial_\nu v_\rho -\frac{2i}{3}v_\mu v_\nu v_\rho)+ \lambda^\alpha \lambda_\alpha \\
&+ 2\partial_3(i\chi_{-}\gamma^m\partial_m\chi_{-}+v_m v^m )].
\end{array}
\end{equation} 



Now, let us consider the finite gauge transformation $g=g_b + \theta^-\psi_-$.  $g^+=g_b^+ + \theta^-\psi_-^+$ with $g g^+=1$. Therefore,
\begin{equation}
\begin{array}{l}
g_b^+ = g_b^{-1}, \\
\psi^+_- = -g_b^+\psi_-g_b^+.
\end{array}
\end{equation}
 Choosing $\Gamma=0$, then $\Gamma^g=\delta_g(\chi_-, v_\mu) = ( -i\psi_- g_b^{-1}, -i\partial_\mu g_b g_b^{-1})$. One can write down the desired boundary action explicitly:
\begin{equation}\label{1nAb}
\begin{array}{ll}
S^{1-}_{nA,b}&=2[-\int d^2x  g_b^{-1}\partial_m g_b g_b^{-1}\partial^m g_b \\& +i\int d^2x\psi_-^+(\gamma^m\partial_m +\gamma^m\partial_m g_b g_b^{-1}) \psi_- \\
& +\frac{2}{3}\int d^3x\epsilon^{\mu\nu\rho} \{g^{-1}_b\partial_\mu g_b g^{-1}_b\partial_\nu g_b g^{-1}_b\partial_\rho g_b\}]
 \\
& =2i\int d^2x d\theta_{-}(g^{-1}\partial_{++} g)(g^{-1}D'_- g)\\&+2i\int d^3x d\theta_{-}[(g^{-1}\partial_{++}g),(g^{-1}\partial_3g)](g^{-1}D'_{-}g).
\end{array}
\end{equation}
The action is invariant under the transformation:
\begin{equation}
\begin{array}{l}
\delta_- g_b = \epsilon^-\psi_-, \\
\delta_- \psi_- = \partial_{--} g_b\epsilon_+.
\end{array}
\end{equation}

Therefore, after we restore the gauge field, the action is consistent
with the gauged $(1,0)$ non-Abelian WZW action $S_{nA,gWZW}^{1-}$ with $g_b\in G$
in \cite{faizal_supersymmetric_2012}: 
\begin{equation}
\label{1gwzw}
\begin{array}{ll}
S^{1-}_{nA,gWZW}& =2i\int d^2x \nabla'_{-}[(g^{-1}\mathcal{D}_{++} g)(g^{-1}\nabla'_- g)]\\
&+2i\int d^3x \nabla'_{-}\{[(g^{-1}\mathcal{D}_{++}g),(g^{-1}\mathcal{D}_3g)](g^{-1}\nabla'_{-}g)\}|_{\theta^-=0}.
\end{array}
\end{equation}

\subsection{\texorpdfstring{$\mathcal{N} = 2$}{N=2} Chern-Simons theory}
We will now apply the results of the previous section to 
$\mathcal{N} = 2$ Chern-Simons theory. The restoration of supersymmetry has previously been considered in the Abelian theory \cite{berman_membranes_2009} and the non-Abelian gauge and supersymmetry invariant superspace actions have been presented in \cite{armoni_defects_2015}. However, the non-Abelian actions have not previously been analyzed in detail.

The action for an $\mathcal{N} =2$ Chern-Simons theory on a manifold 
without boundaries can now be written as \cite{gates_jr._remarks_1992}
\begin{equation}
 S^2 =  \int d^3 x 
\int_0^1 dt d^2\theta d^2\thetab V D^a ( e^{-tV} \bar D_a e^{tV}), \label{csa}
\end{equation}
where the parameter $t$ should not be confused with time $x^0$.
The $\mathcal{N} = 2$ vector superfield $V$ can be expanded into $\mathcal{N} = 1$ component superfields as 
\begin{equation}
V(\theta_1,\theta_2) = A_1(\theta_1) +\theta_2\Gamma_1(\theta_1) -\theta_2^2 (B_1(\theta_1) -D^2_1 A_1(\theta_1)),
\end{equation}
where $A_1$ and $B_1$ are $\mathcal{N} = 1$ real superfields which depend on $\theta_1$, and $\Gamma_1$ is a real spinor superfield, with components summarized by
\begin{equation}
\begin{array}{l}
A_1=a +\theta_1\rho -\theta_1^2 F, \\
 B_1=b +\theta_1\eta -\theta_1^2 k, \\
 \Gamma_1=\chi-\theta_1 M +(\gamma^\mu\theta_1) v_\mu -\theta_1^2[\lambda+ (\gamma^{\mu}\partial_\mu \chi)] .
\end{array}
\end{equation}
For a manifold without boundaries, this action is invariant under $\mathcal{N} =2$ supersymmetry transformations generated by $Q_{a}$ and $\bar Q_{a}$  and also invariant under the following gauge transformations 
\begin{equation}
 e^{V} \to  e^{i \bar{\Sigma} } e^{V} e^{- i \Sigma},
\end{equation}
where $\Sigma$ and $\bar{\Sigma}$ are chiral and anti-chiral superfields respectively.
However, in the case of a boundary, both supersymmetry and gauge symmetry are broken. To preserve half of the supersymmetry, we can modify the action to

\begin{equation}
\label{susyboundary}
S^{\mp\mp}(V)=\int d^{3}x\int_{0}^{1}D_{1\mp}D_{1\pm}{D}_{2\mp}{D}_{2\pm}dt[V D^{a}(e^{-tV}\bar{D}_{a}e^{tV})] \Big\vert_{\theta_1=\theta_2=0}.
\end{equation}

To restore the gauge symmetry, we need to couple other boundary terms. Let us now discuss what boundary terms are required.

\subsubsection*{\texorpdfstring{$\mathcal{N}=2$}{N=2} Abelian Chern-Simons theories}
For simplicity, let us first discuss the Abelian case.
After integrating out one of the spinor coordinates $\theta_2$, the bulk action of the Abelian $\mathcal{N} = 2$ Chern-Simons theory can be rewritten as:
\begin{equation}
\label{2a}
S^2_A=\int d^3x d^2\theta_1 (B_1B_1 + \Gamma_1^\alpha \omega_{1\alpha} +\frac{1}{2}D^\alpha_1(D_{1\alpha}B_1 A_1-B_1D_{1\alpha}A_1)),
\end{equation}
where $\omega_1$ is the super-covariant field strength and $\omega_{1\alpha}=D^{1\beta}D_{1\alpha}\Gamma_{1\beta}$. In the presence of a boundary, we need to consider (\ref{2a}) by adding some boundary terms to recover half of the $\mathcal{N} = 2$ supersymmetry on the boundary. In the following discussion, we mainly focus on the action which preserves $(2,0)$ supersymmetry, although it is also possible to preserve $(1,1)$ supersymmetry. These two cases are related to each other, with $(1, 1)$ supersymmetry resulting from changing the chirality of one of the supercharges in $(2,0)$ supersymmetry. After considering the boundary effect, the action (\ref{susyboundary}) in components \cite{berman_membranes_2009} is:
\begin{equation}
\label{20components}
\begin{array}{ll}
S^{2--}_{A}&=\int d^3x \Big( 2kb+\eta\eta+\lambda\lambda-4\epsilon^{\mu\nu\rho}v_\mu\partial_\nu v_\rho \\
&\quad +2\partial_3( v_m v^m+i\chi_-\gamma^m\partial_m\chi_-+i\rho_-\gamma^m\partial_m\rho_-+\partial_m a\partial^m a-\frac{1}{2}bb) \Big) .
\end{array}
\end{equation}
One may note that here we also introduced some $(2,0)$ supersymmetric invariant action to cancel the gaugino couplings as in $\mathcal{N} = 1$ super Chern-Simons theory and following the same reason below (\ref{restored_gauge}) the gauge transformations of $\rho_+$, $\chi_+$, $M$, $v_{--}$ and $v_3$ lead to the invariance of the action. However, the action is still not gauge invariant because of the gauge transformations of $a$, $v_{++}$, $\rho_-$ and $\chi_-$. As in the $\mathcal{N}=1$ Chern-Simons case, we can also introduce another boundary term:
\begin{equation}
\label{s2_ab}
S^{2--}_{A,b}((e^V)')=S^{2--}_{A}((e^V)^{g,\bar{g}})-S^{2--}_{A}(e^V).
\end{equation}
Here, $(e^V)^{g,\bar{g}}$ denotes the gauge transformation of $(e^V)$ by $g=\exp(i\Lambda')$ and $\bar{g}=\exp(i\Lambdab')$,
\begin{equation}
\label{gaugetrV}
(e^V)^{g,\bar{g}}=\bar{g}\exp(V)g^{-1},
\end{equation}
where $\Lambda'$ and $\Lambdab'$ are $(2,0)$ scalar chiral and anti-chiral superfields, respectively. In Abelian case, this transformation is equivalent to replacing $V$ with $V^{\Lambda',\Lambdab'}=V+i(\Lambdab'-\Lambda')$. We introduce the components of $\Lambda'$ as $\Lambda'=c+\theta^-\psi_- + i\theta^-\thetab^-\partial_{--}c$ and those of $\Lambdab'$ as $\Lambdab'=\bar{c}-\thetab^-\psib_- - i\theta^-\thetab^-\partial_{--}\bar{c}$.  Therefore, we can represent \eqref{s2_ab} with $S_{A,b}^{2--}(V')=S_{A}^{2--}(V^{\Lambda',\Lambdab'})-S_{A}^{2--}(V)$. If we choose the gauge transformation of $g$ and $\bar{g}$ as 
\begin{equation}
g\rightarrow gu^{-1},\quad\bar{g}\rightarrow\bar{g}\bar{{u}}^{-1},\quad e^V \rightarrow (e^V)^{u,\bar{u}}
\end{equation}
the total action $S^{2--}_{A}(e^V)+S^{2--}_{A,b}((e^V)')$ is gauge invariant with $(e^V)^{g,\bar{g}}$ being gauge invariant.

In order to find the explicit form of the boundary term, for simplicity, let us first consider $V=0$. Then $V^{\Lambda',\Lambdab'}=\delta_{g,\bar{g}}V= i(\Lambdab'-\Lambda')$ and the nonvanishing field components are:

\begin{equation}
\label{2tran}
\begin{array}{ll}
\delta_{g,\bar{g}}(a,\rho_-,\chi_-,v_m)&=(\delta_{g,\bar{g}}A_1(a,\rho_-), \delta_{g,\bar{g}}\Gamma_1(\chi_-,v_m))\\
&=(i(\Lambdab'-\Lambda')|_{\theta_{2}^-=0}, iD'_{2-}(\Lambdab'-\Lambda')|_{\theta_{2}^-=0}).
\end{array}
\end{equation}
Since both $A_1$
and $\Gamma_{1}$ are $\mathcal{N}=1$ superfields depending
on $\theta_{1}$, we have the constraint ``$|_{\theta_{2}^-=0}$'' in
the second line. 
For convenience, we
note that we can alternatively use the chirality of
the $(2,0)$ superfields $\Lambda'$ and $\Lambdab'$ to write:
\begin{equation}
\label{change}
\begin{array}{ll}
\delta_{g,\bar{g}}\Gamma_1& =iD'_{2-}(\Lambdab'-\Lambda')|_{\theta_{2}^-=0}\\
&=D'_{1-}(\Lambdab'+\Lambda')|_{\theta_{2}^-=0}.
\end{array}
\end{equation}
Then, $V^{\Lambda',\Lambdab'}$ can be represented by
\begin{equation}
\label{changetran}
\begin{array}{ll}
\delta_{g,\bar{g}}(a,\rho_-,\chi_-,v_m)&=(\delta_{g,\bar{g}}A_1(a,\rho_-), \delta_{g,\bar{g}}\Gamma_1(\chi_-,v_m))\\
&=(i(\Lambdab'-\Lambda')|_{\theta_{2}^-=0},D'_{1-}(\Lambdab'+\Lambda')|_{\theta_{2}^-=0})\\
&=(\Phi_A, D'_{1-}\Phi_B)\\
&=(c_2, \psi_{2-}, \psi_{1-}, \partial_m c_1).
\end{array}
\end{equation}
Here, we define $(1,0)$ scalar superfields $\Phi'_{A}=c_2 + \theta^-_1\psi_{2-}= i(\Lambdab'-\Lambda')|_{\theta_{2}^-=0}$
and $\Phi'_{B}=c_1 + \theta^-_1\psi_{1-}=(\Lambdab'+\Lambda')|_{\theta_{2}^-=0}$. One may note that this
definition with the constraint ``$|_{\theta_{2}^-=0}$'' is not allowed
without the change (\ref{change}), since $D'_{2-}$ can eliminate
$\theta_{2}^-$. 

Therefore, we can write down the required boundary action
\begin{equation}
\label{Abelianboundary}
\begin{array}{ll}
S^{2--}_{A,b}(V') &=S^{2--}_{A}(V^{\Lambda',\Lambdab'})-S^{2--}_{A}(V) \\
&=2\int d^3x\partial_3(\partial_m c_1\partial^m c_1 +i\psi_{1-}\gamma^m\partial_m\psi_{1-} \\
&\quad +i\psi_{2-}\gamma^m\partial_m\psi_{2-} +\partial_m c_2 \partial^m c_2)\\
&=-2i\int d^2x d \theta_{1-} \left( \partial_{++}\Phi'_A D'_{1-}\Phi_A' + \partial_{++}\Phi'_B D'_{1-}\Phi'_B) \right).
\end{array}
\end{equation}
There are some important properties of this action.

First, this action is obviously $(1,0)$ supersymmetry invariant. Actually, the last line of (\ref{Abelianboundary}) is a combination of two standard $(1,0)$ actions. 
Now, we show that this combination is a single $(1,0)$ action. 
For an Abelian gauge group, we can consider the flat target space coordinates to be
 $\tilde{\Phi}'_M=(\Phi'_A, \Phi'_B)$
, with the metric $\eta_{MN}$ being: 
\begin{equation}
\label{metric}
\eta_{MN}=\left(
\begin{array}{ll}
\delta_{AA} & 0\\
0 & \delta_{BB}
\end{array}\right) = \delta_{MN}.
\end{equation}

$M, N$ range in $\{A, B\}$.


We now rename $\theta_{1-}$  as $\tilde{\theta}_{1-}$.
Then, the action (\ref{Abelianboundary}) is 
\begin{equation}
\label{2--ab}
S^{2--}_{A,b}=-2i\int d^2x d \tilde{\theta}_{1-} \eta_{MN}\partial_{++}\tilde{\Phi}'^M \tilde{D}'_{1-}\tilde{\Phi}'^N,
\end{equation}
which is just the standard $(1,0)$ sigma model.

We now represent the $(1,0)$ sigma model as a  $(1,0)$ WZW model \cite{hull_2_1990}, treating the target space $M$ as a group manifold for semi-simple Lie group. Then there are left- and right-invariant vielbeins $L^I_M$ and $R^I_M$, respectively, on the group manifold, they satisfy
\begin{align}
dL^I + \frac{1}{2}f^I_{JK} L^J L^K =0, && dR^I - \frac{1}{2}f^I_{JK} R^J R^K =0,
\end{align}
where $I, J, K$ range in $\{A, B\}$ (the same indices as $M, N$) and are tangent space  indices, i.e. Lie algebra  indices,  $f^I_{JK}$ are the structure constants and $L^I = L^I_M d\tilde{\Phi}'^M$.

When we consider an Abelian group, the situation becomes simple, where
\begin{align}
L^I_M = \delta^I_M, && R^I_M = \delta^I_M, && f^{I}_{JK}=0.
\end{align}
$L^I_M$ is the same as $R^I_M$. 

With respect to the variation of $\tilde{\Phi}'$, one can find the Noether currents
\begin{align}
J^I_{-}=L^I_M D_-\tilde{\Phi}'^M = D_-\tilde{\Phi}'^I, && J^I_{++}=R^I_M\partial_{++}\tilde{\Phi}'^M = \partial_{++}\tilde{\Phi}'^I,
\end{align}
which satisfy 
\begin{align}
\partial_{++} J^I_- =0, && D_- J^I_{++}=0.
\end{align}

Now we use a group $\tilde{g}=\exp i(\tilde{\Phi}'^I t_I)$ instead of $\tilde{\Phi}'$ to specify $M$, and group generators $t_I$ satisfy
\begin{equation}
[t_I, t_J]=f^K_{IJ}t_K.
\end{equation}
Since the generators $t_I$ are two commuting copies of the generators of $H$, in the Abelian case our group field $g$ is valued in $H\times H$. Here, one may note that this $\tilde{g} = \exp(i(\Lambdab'+\Lambda')+i*(i(\Lambdab'-\Lambda'))$ is a boundary field rather than a gauge transformation parameter.

The supersymmetric sigma model can then be represented by a $(1,0)$ WZW model:
\begin{equation}
\label{2AWZW}
S^{2--}_{A,WZW}=2i\Tr\int d^2xd\tilde{\theta}_{1-}(\tilde{g}^{-1}\partial_{++}\tilde{g})(\tilde{g}^{-1}\tilde{D}'_{1-}\tilde{g}), 
\end{equation}
with currents being
\begin{align}
J_{-} = J^I t_I = \tilde{g}^{-1}D_- \tilde{g}, && J_{++} = J^I_{++}t_I = (\partial_{++}\tilde{g})\tilde{g}^{-1}.
\end{align}

One may note that there is no WZ term, since we just considered the Abelian case. One can also restore the gauge fields in the action to obtain a gauged $(1, 0)$ WZW model. 
\begin{equation}
S^{2--}_{A,gWZW}=2i\Tr\int d^2x\tilde{\nabla}'_{1-}[(\tilde{g}^{-1}\mathcal{D}_{++}\tilde{g})(\tilde{g}^{-1}\tilde{\nabla}'_{1-}\tilde{g})]|_{\tilde{\theta}_1^{-} =0}.
\end{equation}

Moreover, the action (\ref{2AWZW}) should be $(2, 0)$ supersymmetry invariant. Spindel \textit{et al.} \cite{spindel_extended_1988, sevrin_extended_1988} proved that a $(1,0)$ sigma model is invariant under $(2,0)$ supersymmetry if the target space admits a complex structure which means the vanishing of Nijenhuis tensor. In the following, we will show that our group manifold endows just such a complex structure. 

We first redefine the coordinates such that an almost complex structure $J^M_N$ is endowed with the standard form:
\begin{equation}
\begin{array}{l}
J^M_N=
\left(
\begin{array}{cc}
0 & -1 \\
1 & 0
\end{array} 
\right).
\end{array}
\end{equation}

It is easy to check that the almost complex structure satisfies the following conditions:
\begin{align}\label{Nijen}
\begin{aligned}
& J^M_N J^N_L = -\delta^M_L,  \quad J^M_N \eta_{ML} + J^M_L \eta_{NM} =0, \\
& \partial_p J^M_N=0,  \quad  \quad  \quad  N_{MN}^L = J^P_M J^L_{[N,P]}-J^P_N J^L_{[M,P]} =0.
\end{aligned} 
\end{align}
The last equation in (\ref{Nijen}) means the vanishing of the Nijenhuis tensor. Of course, for a flat even-dimensional manifold, we can define a global constant complex structure, so this condition is trivially satisfied. Therefore, our action (\ref{2--ab}) is invariant under an $\mathcal{N} = (2,0)$ supersymmetry \cite{spindel_extended_1988, sevrin_extended_1988, hull_2_1990}. 

Now, let us manifestly write down the $(2,0)$ action. Without changing the action, we can introduce a new Grassmann coordinate $\tilde{\theta}_2^-$, and  extend $\tilde{\Phi}'$ to be a $N=(2,0)$ superfield (which means the superfields $\tilde{\Phi}'$, $\Phi'_A$ and $\Phi'_B$ become $(2,0)$ superfields), with a chirality constraint:
\begin{equation}
\tilde{D}'_{-}\tilde{\Phi}'_M= iJ_M^N\tilde{D}'_{-}\tilde{\Phi}'_N,\label{constraint}
\end{equation}
\begin{equation*}
(\tilde{D}'_{1-}-i\tilde{D}'_{2-})\tilde{\Phi}'_A=- i(\tilde{D}'_{1-}-i\tilde{D}'_{2-})\tilde{\Phi}'_B,
\end{equation*}
with $\tilde{\theta}^{-}=(\tilde{\theta}_{1}^{-}+i\tilde{\theta}_{2}^{-})/\sqrt{2}$
and $\tilde{D}'_{-}=(\tilde{D}'_{1-}-i\tilde{D}'_{2-})/\sqrt{2}$. In components, this is
\begin{equation}
\begin{array}{l}
\tilde{D}'_{1-}\Phi_A'=- \tilde{D}'_{2-}\Phi_B',\\
\tilde{D}'_{2-}\Phi_A'= \tilde{D}'_{1-}\Phi_B'.
\end{array}
\end{equation}
One may note that this just means that $\Phi'_{A}=i(\Lambdab'-\Lambda')$
and $\Phi'_{B}=(\Lambdab'+\Lambda')$ without
the constraints $\Lambda'|_{\theta_{2}^-=0}$ and $\Lambdab'|_{\theta_{2}^-=0}$, and $\tilde{\theta}_{2}^{-}$ is $\theta_{2-}$.
Therefore, this chirality constraint naturally exists in our model. 

Based on this constraint, we have 
\begin{equation}
\begin{array}{ll}
S^{2--}_{A,b}&=-2i\int d^2x d\tilde{\theta}_{1-}\partial_{++}\tilde{\Phi}'_M (\tilde{D}'_1)_-\tilde{\Phi}'^M\\
&=-2i\int d^2x d\tilde{\theta}_{2-}\partial_{++}\tilde{\Phi}'_M (\tilde{D}'_2)_-\tilde{\Phi}'^M.
\end{array}
\end{equation}
Therefore, 
\begin{equation}
S^{2--}_{A,b}=-2i\int d^2x [d\tilde{\theta}_-\partial_{++}\tilde{\Phi}'_M \bar{\tilde{D}}'_-\tilde{\Phi}'^M+(c.c)].
\end{equation}
This action is just a $(2,0)$ action. So, we have proved that the boundary action for the $\mathcal{N} = 2$ Chern-Simons action is a $(2,0)$ sigma model \cite{hull_2_1990}.

Furthermore, we can also write this $(2,0)$ model as a $(2,0)$ WZW model, with the group element $\tilde{g}=\exp(i\tilde{\Phi}'^I t_I)$ .
\begin{equation}\label{2Ab}
S^{2--}_{A,WZW}=2i\int d^2x [d\tilde{\theta}_-(\tilde{g}^{-1}\partial_{++}\tilde{g})(\tilde{g}^{-1}\bar{\tilde{D}}'_-\tilde{g})+(c.c)].
\end{equation}

Because of the existence of the complex structure, we can also introduce complex group generators $t_{i}$ from the real group generators $t_I$, by defining
\begin{align}
\begin{aligned}
& t_{i} = \sqrt{\frac{1}{2}}(t_A + i t_{B}), \\
& t_{\bar{i}} = \sqrt{\frac{1}{2}}(t_A - i t_{B}) = (t_i)^*
\end{aligned}
\end{align}
where $A$ and $B$ indicate the two copies of the generators of $H$, but both $A$ and $B$ take the value $i$ in these relations.
It is easy to see that the group element $g$ can be equivalently written as
\begin{equation}
\tilde{g}=\exp(i\tilde{\Phi}'^I t_I) =\exp i(\tilde{\Phi}'^i t_i + \tilde{\Phi}'^{\bar{i}} t_{\bar{i}}),
\end{equation}
where $\tilde{\Phi}'^i=\sqrt{\frac{1}{2}}(\tilde{\Phi}'^A-i\tilde{\Phi}'^B)$ and $\tilde{\Phi}'^{\bar{i}}=\sqrt{\frac{1}{2}}(\tilde{\Phi}'^A+i\tilde{\Phi}'^B$).
Since $t_i$ and $t_{\bar{i}}$ are the generators of complex group $H_C$, we say the group field $\tilde{g}$ is valued in $H_C$ group.  Actually, in Abelian case, $H(n, R)\times H$(n, R) is equal to $H(2n, R)$ which by general redefinitions can transform to $H(n, C)$. Here, $(n, R)$  means that in group $H(n, R)$, there are n parameters valued in $R$.




After we turn on the gauge fields, we have a $(2,0)$ gauged WZW model on the coset space $H_C/H$:
\begin{equation}\label{GAWZW}
S^{2--}_{A,gWZW}=2i\int d^2x \{\tilde{\nabla}'_-[(\tilde{g}^{-1}\mathcal{D}_{++}\tilde{g})(\tilde{g}^{-1}\bar{\tilde{\nabla}}'_-\tilde{g})]+(c.c)\}|_{\tilde{\theta}^- =0}.
\end{equation}

Another interesting property of this boundary action is that the $(2,0)$
WZW model simply reduces to a $(1,0)$ WZW model when we fix the gauge transformation
parameters $\Lambda'|_{\theta_{2}^-=0}$ and $\Lambdab'|_{{\theta}_{2}^-=0}$
to be real or purely imaginary. This can give us a useful hint for
the discussion in the $\mathcal{N}=2$ non-Abelian Chern-Simons case. 

From (\ref{changetran}), it is easy to see that if we restrict $\Lambda'|_{\theta_{2}^-=0}$ and $\Lambdab'|_{\theta_{2}^-=0}$ to be imaginary, i.e.\ $(\Lambdab'+\Lambda')|_{\theta_{2}^-=0}=0$, we only have the gauge transformations for $A_1$, which we denote as $\delta_{gA}A_1$.
\begin{equation}
\delta_{gA}A_1=\delta_{gA}(a,\rho_-)=
( c_2, \psi_{2-}).
\end{equation}
Then, the required boundary action $S^{2--}_{A,bA}$ is a $(1,0)$ model. 
\begin{equation}
\label{2--aba}
\begin{array}{ll}
S^{2--}_{A,bA}&=S^{2--}_{A}(V^{gA})-S^{2--}_{A}(V) \\
&=2\int d^2x (\partial_{++} c_2\partial_{--} c_2+i\psi_{2-}\partial_{++}\psi_{2-})\\
&=-2i\int d^2x d\theta_{1-}\partial_{++}\Phi'_AD'_{1-}\Phi'_A \\
&=2i\int d^2x d\theta_{1-}
(g_A^{-1}\partial_{++}g_A)(g_A^{-1}D'_{1-}g_A), 
\end{array}
\end{equation}
where $g_A=\exp(i\Phi'_A)\in H_A$.

Also, one can fix $(\Lambdab'-\Lambda')|_{\theta_{2}^-=0}=0$, and then only consider the nontrivial gauge transformations of $\Gamma_1$: $\delta_{gB}\Gamma_1=D'_{1-}(\Lambdab'+\Lambda')|_{\theta_{2}^-=0}$. Then the boundary term  $S^{2--}_{A,bB}$ needed is
\begin{equation}
\label{2--abb}
\begin{array}{ll}
S^{2--}_{A,bB}&=S^{2--}_{A}(V^{gB})-S^{2--}_{A}(V)\\
&=2\int d^2x  (\partial_{++} c_1\partial_{--} c_1+i\psi_{1-}\partial_{++}\psi_{1-})\\
&=-2i\int d^2x d\theta_{1-}\partial_{++}\Phi'_B D'_{1-}\Phi'_B \\
&=2i\int d^2x d\theta_{1-}
(g_B^{-1}\partial_{++}g_B)(g_B^{-1}D'_{1-}g_B),
\end{array}
\end{equation}
where $g_B=\exp(i\Phi'_B)\in H_B$.
This is also a $(1,0)$ sigma model.

Actually, one can understand the reduction of the supersymmetry in another way that after we fix the parameters, the complex structure is lost and then only a $(1,0)$ action left.

Remark: there exists an interesting 
correspondence, i.e., the gauge transformation of $\partial_m a$ is just like that of $v_m$, and the gauge transformation of $\rho_-$  is like that of $\chi_-$. The correspondence can be understood in the following way. First, $\Phi'_B$ corresponds to the transformation of $\Gamma_1$: $\delta_{gB}\Gamma_1 = (\psi_{1-}, \partial_m c_1)$. When we only consider the transformations of $\Gamma_1$, $S^{2--}_{A}(V^{gB})$ is equivalent to
\begin{equation}
S^{2--}_{A}(V^{gB})=S^{2--}_{A}(V)+ S^{1-}_A(\Gamma_1^{gB})-S^{1-}_A(\Gamma_1),
\end{equation}
where $S^{1-}_A(\Gamma_1)$ is defined in the same way as that in the $\mathcal{N}=1$ Chern-Simons case. Therefore, the boundary term $S^{2--}_{A,bB}$ is 
\begin{equation}
\begin{array}{ll}
S_{A,bB}^{2--} & =S_{A}^{1-}(\Gamma_{1}^{gB})-S_{A}^{1-}(\Gamma_{1}).\end{array}
\end{equation}
Second, $\Phi'_A$ actually corresponds to the transformation of $\Gamma_2=iD'_{1-}(\Lambdab'-\Lambda')$: $\delta_{gA}\Gamma_2 = (\psi_{2-}, \partial_m c_2)$. We can also rewrite $S^{2--}_{A}(V^{gA})$ as 
\begin{equation}
S^{2--}_{A}(V^{gA})=S^{2--}_{A}(V)+ S^{1-}_A(\Gamma_2^{gA})-S^{1-}_A(\Gamma_2).
\end{equation}
The boundary term $S^{2--}_{A,bA}$ is 
\begin{equation}
\begin{array}{ll}
S_{A,bA}^{2--} & =S_{A}^{1-}(\Gamma_{2}^{gA})-S_{A}^{1-}(\Gamma_{2}).\end{array}
\end{equation}
Both $\Gamma_1$ and $\Gamma_2$ are $\mathcal{N}=1$ spinor superfields and transform in the same way, as do $(\rho_-,\partial_m a)$ and $(\chi_-, v_m)$.
One can note that the boundary action \eqref{2--ab} is the combination of （\eqref{2--aba}） and \eqref{2--abb}. Based on this hint, we now consider the non-Abelian case.

\subsubsection*{\texorpdfstring{$\mathcal{N}=2$}{N=2} non-Abelian Chern-Simons theories preserving \texorpdfstring{$\mathcal{N}=(2,0)$}{N=(2,0)}}
 In the non-Abelian case, the situation becomes complicated without using the Wess-Zumino gauge, or at least the Ivanov gauge. The Chern-Simons action is an infinite series when written in component form, since the commutators do not vanish. Therefore, it is very hard to show an explicit derivation for the boundary theory. Based on our result in the Abelian case, we propose the boundary action contains $S^{2--}_{nA,gWZW}$:
\begin{equation}
\label{2nAb}
\begin{array}{ll}
S^{2--}_{nA,b}&=2i[\int d^2x d\tilde{\theta}_-(\tilde{g}^{-1}\partial_{++}\tilde{g})(\tilde{g}^{-1}\bar{\tilde{D}}'_-\tilde{g})
\\
&\quad +\int d^3x d\tilde{\theta}_-[(\tilde{g}^{-1}\partial_{++}\tilde{g}),(\tilde{g}^{-1}\partial_3\tilde{g})](\tilde{g}^{-1}\bar{\tilde{D}}'_{-}\tilde{g})+(c.c)].
\end{array}
\end{equation}
with group field $\tilde{g}$ valued in $G_C$ group.  When we define $\tilde{g}=\tilde{g}_b + \tilde{\theta}^-\tilde{\psi}_-$, in components, the action is:
\begin{equation}
\begin{array}{ll}
S^{2--}_{nA,b}&=2[-\int d^2x (\tilde{g}^{-1}_b\partial_{++}\tilde{g}_b)(\tilde{g}^{-1}_b\partial_{--}\tilde{g}_b)
\\
&\quad +i\int d^2x \tilde{\psi}_-^+(\gamma^m\partial_m +\gamma^m\partial_m \tilde{g}_b \tilde{g}_b^{-1})\tilde{\psi}_- \\
& \quad +\frac{2}{3}\int d^3x\epsilon^{\mu\nu\rho}(\tilde{g}^{-1}_b\partial_\mu\tilde{g}_b)(\tilde{g}^{-1}_b\partial_\nu\tilde{g}_b)(\tilde{g}^{-1}_b \partial_\rho\tilde{g}_b)].
\end{array}
\end{equation}

Now, let us put our effort to justify this action is the right result. 

First, when we consider the gauge group $G$ being an Abelian group, we see that the non-Abelian WZW action \eqref{2nAb} reduces to the Abelian action \eqref{2Ab}. Therefore, the group elements should become group elements valued in $H\times H$ or $H_C$.

Second, we know that the bulk theory of the $\mathcal{N}=2$ Chern-Simons action can be written as an $\mathcal{N}=1$ Chern-Simons action with a term of an auxiliary field: 
\begin{equation}
S^{2--}_{nA,b}=\int d^3 x \int d^2\theta_1 (\Gamma_1\Omega_1 +B_1^2).
\end{equation}
If we only preserve $\mathcal{N}=(1,0)$ supersymmetry on the boundary, we only need to restore the gauge symmetry and supersymmetry of the $\mathcal{N}=1$ Chern-Simons theory which we know is the $\mathcal{N}=(1,0)$ non-Abelian  action \eqref{1nAb} with group field valued in group $G$. This can be given by the $\mathcal{N}=(2,0)$ non-Abelian action \eqref{2nAb} by restricting $\tilde{g} \in G_C$ to the subgroup $G$, i.e.\ $\tilde{g} \in G$. 

To satisfy the above two requirements, there are two possible (minimal) choices: $(2,0)$ non-Abelian WZW model with group elements valued in $G\times G$ or $G_C$. One may note that in the Abelian case, we introduced two groups $H\times H$ and $H_C$, which are equivalent in that case, since all generators commute with each other. There are two types of gauge transformations with parameters $\exp[i(\Lambdab' +\Lambda')]$ and $\exp[i \times (i(\Lambdab' -\Lambda'))]$.
However, in the non-Abelian case, $G\times G$ is not equal to $G_C$. So the question is which one is the right choice. The reason we choose $G_C$ is shown as follows\footnote{We thank the anonymous referee for comments to clarify this issue.}. First of all, it is obvious that the superfield transformation rule involves $G_C$ since $\Sigma$ is complex. Second, if the right group is $G\times G$, there should be two real subgroups $G$. However, in the non-Abelian case, when restricting to the imaginary part by setting $\Lambda' = - \Lambdab'$, i.e., $g^{-1} = \bar{g} = M =\exp [i \times (i(\Lambdab' -\Lambda'))]$, the gauge transformation rule is 
\begin{align}
& (\exp V)^M = M\exp V M, \\ 
& ((\exp V)^M)^N= NM(e^V)M^{-1}N^{-1} = (NM)(e^V)(MN).
\end{align}
In general, $NM$ does not equal $MN$, so the imaginary part does not give rise to a closed subgroup of $G_C$, unless we are in the Abelian case.  While restricting $\Lambda' = \Lambdab'$ and then $g = \bar{g} = \exp [i(\Lambdab' +\Lambda')]$, we get the standard gauge transformation rule as
\begin{align}
& (\exp V)^M = M\exp V M^{-1}, \\ 
& ((\exp V)^M)^N= NM(e^V)M^{-1}N^{-1} = (NM)(e^V)(NM)^{-1}.
\end{align}
This forms a subgroup $G$ of $G_C$. 
This is consistent with the fact that a non-Abelian $G_C$ has only one real subgroup $G$ by restricting the group elements real. Therefore, we choose $G_C$ group instead of $G\times G$.

After we turn on the gauge fields, we obtain the gauged $(2,0)$ WZW model:
\begin{equation}
\label{2nAb_Gauge}
\begin{array}{ll}
S^{2--}_{nA,b}&=2i[\int d^2x d\tilde{\nabla}_-(\tilde{g}^{-1}\partial_{++}\tilde{g})(\tilde{g}^{-1}\bar{\tilde{\nabla}}'_-\tilde{g})
\\
&\quad +\int d^3x d\tilde{\nabla}_-[(\tilde{g}^{-1}\partial_{++}\tilde{g}),(\tilde{g}^{-1}\partial_3\tilde{g})](\tilde{g}^{-1}\bar{\tilde{\nabla}}'_{-}\tilde{g})+(c.c)].
\end{array}
\end{equation}

\subsubsection*{\texorpdfstring{$\mathcal{N}=2$}{N=2} non-Abelian Chern-Simons theories preserving \texorpdfstring{$\mathcal{N}=(1,1)$}{N=(1,1)}}

One should also obtain a $(1,1)$ WZW model  when preserving $\epsilon^{1+}Q_{1+}$ and $\epsilon^{2-}Q_{2-}$ (or $\epsilon^{1-}Q_{1-}$ and $\epsilon^{2+}Q_{2+}$)
on the boundary simultaneously. 
Then, to preserve the gauge symmetry, considering the
transformation of $V$ by $\hat{g}=\exp(i\hat{\Phi}')$ which
is a $(1,1)$ superfield belonging to $G$ , we also propose the boundary
action being a $(1,1)$ WZW model \cite{vecchia_supersymmetric_1985}
\begin{equation}
\begin{array}{ll}
S_{nA,b}^{2+-} & =\int d^{2}xd^{2}\hat{\theta}\bar{\hat{D}}'\hat{g}^+\hat{D}'\hat{g}\\
 & +2\int d^{3}xd^{2}\hat{\theta}\hat{g}{}^{-1}\partial_{3}\hat{g}\bar{\hat{D}}'\hat{g}^{+}\gamma_{3}\hat{D}'\hat{g},
\end{array}\label{11naf-1}
\end{equation}
where $\hat{D}'=\left(\begin{array}{c}
\hat{D}'_{+}\\
\hat{D}'_{-}
\end{array}\right)$ and $\hat{D}'_\pm = D'_\pm$. (Here we use $\hat{}$ to denote fields in this case to hopefully avoid confusion with the fields discussed
in the $(2,0)$ case.) 
We can redefine the components of $\hat{g}=\hat{g}_b + \hat{\theta}\hat{\psi}-\hat{\theta}^2 \hat{f}$ and $\hat{g}^+=\hat{g}_b^+ + \hat{\theta}\hat{\psi}^+ -\hat{\theta}^2 \hat{f}^+$. $\hat{g}$ and $\hat{g}^+$ satisfy the constraint $g^+ g=1$:
\begin{equation}
\begin{array}{l}
\hat{g}^+_b = \hat{g}^{-1}_b,\\
\hat{\psi}^+= -\hat{g}^+_b\hat{\psi}\hat{g}^+_b,\\
\hat{f}^+ = -\hat{g}^+_b \hat{f}\hat{g}^+_b-\hat{\psi}^+\hat{\psi}\hat{g}^+_b.
\end{array}
\end{equation}

The action in components can be written as
\begin{equation}
\begin{array}{ll}
S_{nA,b}^{2+-} &=
2(\int d^2x \partial_{++}g^+\partial_{--}g +\int d^3x \frac{2}{3} \epsilon^{\mu\nu\rho}\partial_\mu g g^+\partial_\nu g g^+\partial_\rho g g^+ )\\
&\quad +2\int d^2x(\psi^+\gamma_3\gamma^m\partial_m g g^+\psi+\psi^+\gamma^m\partial_m \psi) \\
&\quad +  2\int d^2x [
-f^+ f + \frac{1}{2}  \psi^+\gamma_3\psi (g^+f-f^+g)]
\end{array}
\end{equation}

By construction, the (1,1) action is invariant under the transformation:
\begin{equation}
\begin{array}{l}
\delta \hat{g}_b = \epsilon^-\hat{\psi}_- +\epsilon^+\hat{\psi}_+, \\ \delta\hat{\psi} = -\epsilon \hat{f} + (\gamma^\mu\epsilon)\partial_\mu \hat{g}_b, \\
\delta \hat{f} = -\epsilon(\gamma^\mu\partial_\mu)\hat{\psi}.
\end{array}
\end{equation}
There is no kinetic term for the auxiliary field $\hat{f}$. Therefore, we can integrate it out by replacing it with its equation of motion:
\begin{equation}
\begin{array}{ll}
\hat{f}=\hat{\psi}\hat{g}_b^{-1} P_+\hat{\psi}.
\end{array}
\end{equation}
Using this together with the constraint $\hat{g}^+ \hat{g}=1$, we can eliminate the auxiliary field 
so that we are left with the following expression
\begin{equation}
\begin{array}{ll}
S_{nA,b}^{2+-} &=2\int d^2x \partial_m \hat{g}_b^+\partial^m \hat{g}_b  \\
& +\frac{4}{3}\int d^3x\epsilon^{\mu\nu\rho}\Tr\{\hat{g}^{-1}_b\partial_\mu \hat{g}_b \hat{g}^{-1}_b\partial_\nu \hat{g}_b \hat{g}^{-1}_b\partial_\rho \hat{g}_b\}
\\& +2 \int d^2x\Tr {\hat{\psi}}^+(\gamma^m\partial_m +\gamma_3\gamma^m\partial_m \hat{g}_b \hat{g}_b^{-1} )\hat{\psi},
\end{array}
\end{equation}
where we have used the identity
\begin{equation}
\Tr\{(\hat{\psi}^+\hat{\psi})^2 +(\hat{\psi}^+\gamma^3\hat{\psi})^2\}=0,
\end{equation}
which is valid for Majorana fermions. Then, we know that the $(1,1)$ action can be written as a free fermion action. The supersymmetry transformation is 
\begin{equation}
\begin{array}{l}
\delta \hat{g}_b = \epsilon^-\hat{\psi}_- +\epsilon^+\hat{\psi}_+, \\ \delta\hat{\psi} = -\epsilon (\hat{\psi}\hat{g}_b^{-1}P_{+}\hat{\psi}) + (\gamma^\mu\epsilon)\partial_\mu \hat{g}_b .
\end{array}
\end{equation}
We can also check the action by reducing it to a $(1,0)$ WZW model or a  $(0,1)$ WZW
model. When we choose to preserve only $\epsilon^{2-}Q_{2-}$ or $\epsilon^{1+}Q_{1+}$:
when the gauge transformation parameter is fixed to be a $(1,0)$
scalar superfield $\hat{g}_{A}=\hat{g}_{\theta_{1+} = 0}$
then the required boundary action \eqref{11naf-1} becomes a $(1, 0)$
WZW model 
\begin{equation}
\begin{array}{ll}
S_{nA,bA}^{2+-} & =2i\int d^{2}xd\hat{\theta}_{-}({\hat{g}}_{A}^{-1}\partial_{++}{\hat{g}}_{A})({\hat{g}}_{A}^{-1}\hat{D}'_{-}{\hat{g}}_{A})\\
 & +2i\int d^{3}xd\hat{\theta}_{-}[({\hat{g}}_{A}^{-1}\partial_{++}{\hat{g}}_{A}),(\hat{g}_{A}^{-1}\partial_{3}{\hat{g}}_{A})]({\hat{g}}_{A}^{-1}\hat{D}'_{-}{\hat{g}}_{A});
\end{array}\label{nA,bA}
\end{equation}
and, when the gauge
transformation parameter is fixed to be a $(0,1)$ scalar superfield
$\hat{g}_{B}=\hat{g}_{\theta_{2-}  =0}$, the action \eqref{11naf-1}
becomes a $(0, 1)$ WZW model 
\begin{equation}
\begin{array}{ll}
S_{nA,bB}^{2+-} & =-2i\int d^{2}xd\hat{\theta}_{+}({\hat{g}}_{B}^+\partial_{--}{\hat{g}}_{B})({\hat{g}}_{B}^+\hat{D}'_{+}{\hat{g}}_{B})\\
 & +2i\int d^{3}xd\hat{\theta}_{+}[({\hat{g}}_{B}^+\partial_{--}{\hat{g}}_{B}),({\hat{g}}_{B}^+\partial_{3}{\hat{g}}_{B})]({\hat{g}}_{B}^+\hat{D}'_{+}{\hat{g}}_{B}).
\end{array}\label{nA,bB}
\end{equation}
Actually, sharing the same bosonic terms, the combination
of (\ref{nA,bB}) and (\ref{nA,bA}) can be the $(1,1)$ WZW action
\eqref{11naf-1}  whereby
the $(1,1)$ action can be written as a WZW action of free fermions
\cite{vecchia_supersymmetric_1985,papadopoulos_solitons_1995}.

After we turn on the gauge fields, we obtain the gauged $(1,1)$ WZW
model: 
\begin{equation}
\begin{array}{ll}
S_{nA,b}^{2+-} & =-\int d^{2}x\hat{\nabla}'^{2}(\hat{g}^+\bar{\hat{\nabla}}'\hat{g}\hat{g}^+\hat{\nabla}'\hat{g}) \\& -2\int d^{3}x\hat{\nabla}'^{2}(\hat{g}^+\partial_{3}\hat{g}\hat{g}^+\bar{\hat{\nabla}}'\hat{g}\hat{g}^+\gamma_{3}\hat{\nabla}'\hat{g})|_{\hat{\theta}=\hat{\thetab}=0}.\end{array}
\end{equation}
One difference with the $(2,0)$ WZW model is that the $(1,1)$ WZW
model does not require the existence of a complex structure.

\section{Conclusion}
In this paper we first considered the addition of a boundary in $\mathcal{N} = 2$ theories, and analyzed the minimal additional boundary action which must be included to restore half the supersymmetry. We then applied this formalism to $\mathcal{N} = 2$ Chern-Simons theories. As is well known, this has the extra complication that the presence of a boundary also breaks the gauge symmetry. However, we showed that restoring the full gauge symmetry is possible, and that doing so leads to a supersymmetric gauged WZW model on the boundary. While performing this analysis in $\mathcal{N} = 2$ superspace proved too technically challenging, we were able to work in $\mathcal{N} = 1$ to derive the result. We could explicitly show that in the case of boundary $\mathcal{N} = (2, 0)$ supersymmetry, the manifestly $\mathcal{N} = (1, 0)$ sigma model possessed a complex structure so that the required conditions were met for enhancement of supersymmetry.

It is possible to use these results to analyze
ABJM theory in the presence of a boundary. It is important to perform such an analysis, as the ABJM theory describes multiple M2-branes and 
 multiple
M2-branes  can end  on M5-branes, M9-branes or gravitational waves. 
By writing the ABJM theory in $\mathcal{N} = 2$ superspace, it is possible to modify the original Lagrangian by introducing boundary terms which preserve half 
the supersymmetry.  
Furthermore, the  
matter Lagrangian will be  gauge invariant by itself in the presence of a boundary, while 
the gauge sector of this theory can be made gauge invariant by adding new boundary degrees of freedom as presented in this article. These new boundary degrees of freedom will include WZW models with either
 $\mathcal{N} = (2, 0)$  or $\mathcal{N} = (1, 1)$ supersymmetry. 
Thus, it is possible to obtain the full 
gauge invariant Lagrangian for the ABJM theory, which preserves half the manifest supersymmetry of the original theory in $\mathcal{N} = 2$ superspace formalism. 
However, we are also free to include additional gauge- and supersymmetry-invariant boundary terms. Such terms are in part determined by the manifest supersymmetry, see e.g.\ \cite{berman_membranes_2009} for
a discussion of the boundary potential for ABJM theory.
The full $\mathcal{N} = 6$ superconformal 
invariance should restrict the boundary potential to be a specific quartic potential.
It would be interesting to perform a complete analysis for the non-Abelian ABJM theory in $\mathcal{N} =2$ superspace formalism using additional R-symmetry constraints to derive the enhanced supersymmetric boundary action. 
It may be noted that the ABJM theory has also been formulated in $\mathcal{N} =3$ harmonic superspace \cite{buchbinder_abjm_2009}. It would thus also be interesting 
to analyze the ABJM theory in the presence of a boundary, in $\mathcal{N} =3$ harmonic superspace.

\section*{Acknowledgements}

The author Q.Z. would like to thank Junya Yagi for a lot of illuminating and helpful discussions.
D.J.S. is supported in part by the STFC Consolidated Grant ST/L000407/1. The work of Y.L., M.-C.T. and Q.Z. is supported by NUS Tier 1 FRC Grant R-144-000-316-112.

\appendix
\section{Appendix}
\subsection{Conventions} \label{convention}
In this paper, we assume Lorentzian $(-++)$ signature. In our conventions,
\begin{equation}
\begin{array}{lll}
\theta^\alpha= C^{\alpha\beta}\theta_\beta, & \theta_\alpha=\theta^\beta C_{\beta\alpha}, & \theta^2 =\frac{1}{2}\theta^\alpha\theta_\alpha = -\frac{1}{2}\theta_\beta C^{\beta_\alpha}\theta_\alpha. 
\end{array}
\end{equation}
Here, we consider 
\begin{equation}
C_{\alpha \beta} = -C^{\alpha\beta}= \left( 
\begin{array}{ll}
0 & -i \\
i & 0
\end{array} \right).
\end{equation}

\begin{equation}
\begin{array}{lll}
\gamma^{0\beta }_\alpha= \left( 
\begin{array}{ll}
0 &1 \\
-1 & 0
\end{array} \right), & 
\gamma^{1\beta }_\alpha = \left( 
\begin{array}{ll}
0 &1 \\
1 & 0
\end{array} \right), &
\gamma^{3\beta }_\alpha= \left( 
\begin{array}{ll}
1 & 0 \\
0 & -1
\end{array} \right).
\end{array}
\end{equation}

\begin{equation}
\begin{array}{rl}
\gamma^\mu_{\alpha\beta} \equiv \gamma^{\mu\gamma}_\alpha  C_{\gamma\beta} = \gamma^\mu_{\beta\alpha}, & \gamma^\mu\gamma^\nu = \eta^{\mu\nu} + \epsilon^{\mu\nu\rho}\gamma_\rho, \\
  P_\pm = \frac{1}{2}(1\pm \gamma_3), & \epsilon^{013}=1.
\end{array}
\end{equation}

Differentiation and integration is summarized by 
\begin{equation}
\begin{array}{lll}
\partial_\alpha\theta^\beta = \delta^\beta_\alpha, &\int d\theta_\alpha\theta^\beta = \delta_\alpha^\beta, & \int d^2\theta \theta^2 =-1.
\end{array}
\end{equation}
The supercharges and covariant derivative are
\begin{equation}
\begin{array}{ll}
D_\alpha = \partial_\alpha +(\gamma^\mu\theta)_\alpha\partial_\mu, &
Q_\alpha = \partial_\alpha - (\gamma^\mu\theta)_\alpha\partial_\mu, \\ 
D^\alpha = -\partial^\alpha - (\theta\gamma^\mu)^\alpha \partial_\mu. 
\end{array}
\end{equation}
We obtain the following algebra
\begin{equation}
\begin{array}{lll}
\{Q_\alpha, Q_\beta\}=2\gamma^\mu_{\alpha\beta}\partial_\mu, & \{Q_\alpha, D_\beta\}=0, & \{D_\alpha, D_\beta\}=-2\gamma^\mu_{\alpha\beta}\partial_\mu.
\end{array}
\end{equation}

\subsection{Decomposition: \texorpdfstring{$\mathcal{N}=1 \rightarrow \mathcal{N}=(1,0)$}{N=1 --> N=(1,0)} } \label{appendix2}
We can first define the boundary supercharges
\begin{equation}
\begin{array}{l}
Q'_{\mp}=\partial_\mp -(\gamma^m\theta)_\mp\partial_{m}.\\
Q'_{\mp}=\partial_\mp -(\gamma^m\theta)_\mp{\partial_m}
\end{array}
\end{equation}
Then, we have
\begin{equation}
\begin{array}{ll}
Q'_-=Q_- -\theta_-\partial_3, & Q'_+ = Q_+ +\theta_+\partial_3. 
\end{array}
\end{equation}
The two operators $Q'_-$ and $Q_-$ are related as follows
\begin{equation}
\begin{array}{l}
Q'_- = Q_- -\theta_-\partial_3 = \exp(\theta^-\theta_-\partial_3)Q_-\exp(-\theta^-\theta_-\partial_3)\\
Q'_+ = Q_+ +\theta_+\partial_3 = \exp(-\theta^-\theta_-\partial_3)Q_+\exp(\theta^-\theta_-\partial_3)
\end{array}
\end{equation}
Therefore, we write 
\begin{equation}
\begin{array}{ll}
\Phi & = \exp(-\theta^-\theta_-\partial_3)(\hat{A}(\theta^-)+\theta^+\hat{A}_+(\theta^-)).
\end{array}
\end{equation}
So, one can write down the half supersymmetric multiplets from a fully supersymmetric one. Let us first discuss the scalar multiplet
\begin{equation}
\begin{array}{ll}
A &= a +\theta\psi -\theta^2 f \\
& = \exp(-\theta^-\theta_-\partial_3)(\hat{A}(\theta^-) +\theta^+ \hat{A}_+(\theta_+)),
\end{array}
\end{equation}
where hatted objects are now boundary superfields (1+1 dimension) whose supersymmetry is generated by $\epsilon^-Q'_- \equiv \epsilon^-(\partial_- -(\gamma^m\theta)_-\partial_m)$. 
Then one can obtain the half multiplets:
\begin{equation}
\begin{array}{ll}
\hat{A}= a +\theta^-\psi_-, & \hat{A}_+ = \psi_+ -\theta^-(f -\partial_3 a).
\end{array}
\end{equation}

The spinor multiplet can be decomposed in a similar way: 
\begin{equation}
\begin{array}{ll}
\Gamma_{\mp}& =\chi_\mp -\theta_\mp M +(\gamma^\mu\theta)_\mp v_\mu -\theta^2[\lambda +\gamma^\mu\partial_\mu \chi]_\mp \\
\Gamma_- & = \exp(-\theta^-\theta_-\partial_3)(\hat{\Gamma}_{-}(\theta^-) -\theta_-\hat{\Sigma}^-(\theta^-)] \\
\Gamma_+ & =\exp (-\theta^-\theta_-\partial_3)[\hat{\Gamma}_+(\theta_+) +(\gamma^m\theta)_+\hat{\Sigma}^+_m(\theta_+)]
\end{array}
\end{equation} 
\begin{equation}\label{boundary_super}
\begin{array}{rl}
\hat{\Gamma}_- &= \chi_- +(\gamma^m\theta)_-v_m, \\
\hat{\Sigma}^- &= M +v_3 -\theta^-[\lambda_- -2\partial_3\chi_-+ (\gamma^m\partial_m\chi)_-], \\
\hat{\Gamma}_+ &= \chi_+ +\theta_+(-M +v_3), \\
(\gamma^m\theta)_+\hat{\Sigma}^+_m & = (\gamma^m\theta)_+[ v_m + \theta^-(\frac{1}{2}\gamma_m\lambda +\partial_m\chi)_-].
\end{array}
\end{equation}

\newpage 

\bibliographystyle{elsarticle-num}
\bibliography{references_zhaoqin_final_no_URL_DOI}

\end{document}